\documentclass[aps,prr,reprint,superscriptaddress]{revtex4-2}
\usepackage{marginnote}
\usepackage{color,graphicx,hyperref}
\usepackage{epsfig,amsmath,amssymb,calc,verbatim,physics,ulem}
\usepackage{bm} 

\newcommand{\eff}{\mbox{\scriptsize eff}}
\newcommand{\hex}{{\mbox{\scriptsize Hex}}}

 \newcommand{\pdif}[2]{\frac{\partial #1}{\partial #2 } }
\newcommand{\hate}{\hat{\bm e}}
\newcommand{\bq}{{\bm q}}
\newcommand{\br}{{\bm r}}
\newcommand{\bG}{{\bm G}}
\newcommand{\bbeta}{{\bm \eta}}

\begin{document}
\title{Crystallization of Chiral Active Brownian Particles at Low Densities}
\affiliation{Department of Physics, Nagoya University, Nagoya 464-8602, Japan}
\author{Kangeun Jeong}
\email{jeong@r.phys.nagoya-u.ac.jp}
\author{Yuta Kuroda}
\affiliation{Department of Physics, Nagoya University, Nagoya 464-8602, Japan}
\author{Yuki Asatani}
\affiliation{Department of Physics, Nagoya University, Nagoya 464-8602, Japan}
\author{Takeshi Kawasaki}
\affiliation{Department of Physics, Nagoya University, Nagoya 464-8602, Japan}
\affiliation{D3 Center, Department of Physics, Osaka University, Toyonaka, Osaka 560-0043, Japan}
\author{Kunimasa Miyazaki}
\email{miyazaki@r.phys.nagoya-u.ac.jp}
\affiliation{Department of Physics, Nagoya University, Nagoya 464-8602, Japan}
\date{\today}

\begin{abstract}
Chiral active matter is a variant of active matter systems in which the motion
 of the constituent particles violates mirror symmetry. In this letter,
 we simulate two-dimensional chiral Active Brownian Particles, the simplest chiral model in which each particle undergoes 
 circular motion, and show that the system crystallizes at low densities well below the melting point
 of the equilibrium counterpart. 
Crystallization is only possible if the orbital radius is long enough to
 align the circulating particles, but short enough for neighboring
 particles to avoid collisions. 
Of course, the system
 must be driven sufficiently far from equilibrium, since chirality
 cannot affect thermodynamic properties in classical
 equilibrium systems. The fluid-crystal phase diagram shows a
 re-entrant melting transition as a function of the radius of the
 circles. We show that at least one of the two transitions follows the
same two-step melting scenario as in equilibrium systems. 
\end{abstract}
\maketitle

The macroscopic properties of classical systems in 
equilibrium are known to be unaffected by chirality 
(or broken mirror symmetry) in 
the underlying microscopic equations of motion 
~\cite{landaustaphys1,Kubo1990statphys-recitation}. 
For example, a magnetic field that induces chiral motion of point
charges cannot induce macroscopic magnetization, according to the
Bohr-van Leeuwen theorem~\cite{Bohr1911,vanLeeuwen1919}. 
This is not the case when the system is taken out of equilibrium,
where it is expected that the interplay of chirality and violation
of detailed balance will lead to new phenomena not found in equilibrium systems. 
Active matter, due to its intrinsic nonequilibrium nature, offers an ideal testbed for studying such phenomena~\cite{Liebchen2022epl}.
Chiral motions are abundant in biological systems.
Many living systems, such as bacteria, 
cells, and algae, perform circular and spinning motions~\cite{Lauga2006bj,Riedel2005sci,Huang2021pnas}. 
Chiral active matter systems can also be easily designed
and realized using synthetic particles, such as asymmetric colloidal particles and Quincke rollers~\cite{Kummel2013prl,Zhang2020natc}.
Over the past decade, studies on chiral active matter have 
revealed novel and rich collective behaviors not observed in non-chiral systems. 
These include vortex formation in flocking systems~\cite{Zhang2020natc},
non-reciprocal transport~\cite{Banerjee2017natc}, and even the crystallization of spinning particles~\cite{Petroff2015prl,Tan2022nat,Bililign2022natp, Caporusso2024prl}.
However, the microscopic mechanisms underlying these exotic phenomena remain largely elusive. 
In particular, it is desirable to understand how the chirality of
the microscopic equations affects the fluid-solid phase transition and
the glass (or jamming) transitions~\cite{Debets2023,Arora2024}. 

In this letter, we focus on the crystallization of a chiral active
matter model consisting of circularly orbiting particles. To this end,
we use the simplest model of this kind, chiral Active Brownian
Particles (cABP).
The cABP model is a chiral variant of the ABP model~\cite{Ma2017sm}. 
The equation of motion of the $i$-th cABP is assumed to be governed by an overdamped Langevin equation:
\begin{equation}
 \begin{aligned}
 \dot{\br}_i=- \frac{1}{\zeta} \pdif{U}{\br_i} + v_0 \hate_i, 
\label{1}
 \end{aligned}
\end{equation}
where $\zeta$ is the drag coefficient and $U$ is the interaction
potential. 
$\hat{\bm e}_i$ is the orientation of the self-propulsion force, which satisfies
\begin{equation}
\dot{\hat{\bm e}}_i
=\left(\Omega \hat{\bm z}+ \sqrt{\frac{2}{\tau_p}}~\bbeta_i  \right)
\times \hate_i. 
\label{2}
\end{equation}
Here, $\bbeta_i$ represents Gaussian white noise with zero mean, which 
satisfies
$\ev{\eta_{i,\alpha}(t) \eta_{j,\beta}(t')} 
= 
\delta_{ij}\delta_{\alpha\beta} \delta(t - t')$, 
where $\alpha$ and $\beta$ represent the coordinates. 
$v_0$ and $\tau_p$ denote the noise strength and its persistence time, respectively.
$\Omega$ is the angular frequency with respect to the rotation axis $\hat{\bm z}$
that causes the particles to move in a circular motion with a radius $R=v_0/\Omega$. 
The two key parameters of the model are $\tau_p$,
which controls the distance from equilibrium and 
$R$, which 
characterizes the extent of
chiral motion. 
When $\tau_p$ is very small, the fluctuation-dissipation theorem is recovered, 
and thus the equation becomes that of equilibrium Brownian motion, even when $\Omega$ is finite. 
Recently, the two-dimensional system in the large $\tau_p$ limit has been studied extensively 
at low densities by Lei \textit{et al.}~\cite{Lei2019sa}. 
They discovered that the system undergoes an absorbing phase transition from
a diffusive ergodic state to a quiescent nonergodic state.  
The most striking feature of cABP in the absence of stochastic noise
($\tau_p=\infty$) is hyperuniformity (HU): strong suppression 
of density fluctuations over long length scales~\cite{Torquato2018pr}. 
According to Ref.~\cite{Lei2019sa}, HU has been observed not only in the vicinity  of
the absorbing phase transition~\cite{Henkel2008book}, but also throughout an entire 
active phase. 
Kuroda {\it et al.} have investigated the fate of this two-dimensional noiseless cABP  at high
densities and found that the system crystallizes~\cite{Kuroda2025prr}. 
Surprisingly, this crystalline state is characterized by true long-range
translational order, which is strictly forbidden in equilibrium systems in two dimensions 
by the Mermin-Wagner theorem~\cite{Mermin1968pr}. 
The authors of Ref.~\cite{Kuroda2025prr} argued that  the HU of cABP is responsible for
the ultra-stability of the crystalline phase. 
This stability persists as long as $\Omega \neq 0$, where HU survives. 
However, this argument applies only for the noiseless  limit, \textit{i.e.}, $\tau_p= \infty$.

Here, we consider the case in which $\tau_p$ is finite and the density
is relatively low. 
We numerically simulate the cABP system for a wide range of 
$\Omega$, $\tau_p$, and densities.
We find that the system crystallizes at densities well
below the melting point of equilibrium two-dimensional systems.  
The mechanism of crystal formation is deceptively simple. 
The particles undergo 
circular motion with a radius $R=v_0/\Omega$, which  
increases their effective radius. This reduces the available free volume and promotes crystallization. 
Crystallization occurs when the effective packing fraction defined by the
effective radius becomes sufficiently high.
However, if $R$ becomes too large, strong collisions between particles
distort their circular orbits, render their trajectories chaotic, and
eventually destroy the crystalline structure.
The persistence time, $\tau_p$, should also be within the appropriate range. 
If $\tau_p$ is infinitesimally small, 
the noise in Eq.~(\ref{1}) becomes
Markovian, detailed balance is restored, and thus chirality
cannot induce crystallization even when $R$ is finite. 
When $\tau_p$ is infinite,  the noise in Eq.~(\ref{2}) disappears, and
thus the system enters an absorbing state, in which each particle
resides at a fixed point without interacting~\cite{Lei2019sa}. 
The crystal that we observe showcases order-by-disorder, as finite noise is essential~\cite{Hanai2024}. 
Additionally, we found that crystallization follows the two-step melting
transition scenario as in the 
Kosterlitz-Thouless-Halperin-Nelson-Young (KTHNY)
theory~\cite{Kardar2007book2}. 

We simulate one-component cABP systems in two dimensions by solving 
Eqs.~(\ref{1}) and (\ref{2}). 
For the inter-particle potential $U = \sum_{i>j}\phi(r_{ij})$,  where
$r_{ij}$ is the distance between the $i$-th and $j$-th particles, 
we use the Weeks-Chandler-Andersen (WCA) potential
\begin{equation} 
\phi(r) = 4\epsilon\left\{ \left(\frac{\sigma}{{r}}\right)^{12} -
    \left(\frac{\sigma}{{r}}\right)^6 + \frac{1}{4} \right\} \Theta\left(2^{1/6}\sigma - {r}\right),  
\label{WCA}
\end{equation}
where $\sigma$ is the particle diameter, and $\Theta(x)$ is the Heaviside step function. 
The interaction potential has a cutoff length of $ 2^{1/6}\sigma$,
resulting in purely repulsive forces between particles.
We choose $\sigma$ and $\sigma/v_0$ as units of length and time, respectively. 
The energy ratio $\epsilon/v_0\zeta\sigma$ is set to 1/24~\cite{Lei2019sa}.
The important control parameters in the present study are
the orbital radius $R = v_0/\Omega$, 
the persistence time $\tau_p$, and the density (or packing fraction) 
$\varphi = N\pi\sigma^2/(4L_xL_y)$, where $N$ is the total number of particles, 
and $L_x$ and $L_y$ are the side lengths of the simulation box. 

To accommodate the hexagonal lattice under periodic boundary conditions, 
the simulation box is usually designed with an aspect ratio of $L_x : L_y = 7 : 4\sqrt{3}$~\cite{Kuroda2025prr}.
Most simulations are carried out for the system size of $N=9\,464$. 
Larger systems of $N = 40\,824$ are used to compute correlation
functions with high precision near the transition point. 
Further details of the simulation setup can be found in the Supplementary Material~\cite{SM}. 

\begin{figure}[tb]
\includegraphics[width=1.0\columnwidth,angle=-0]{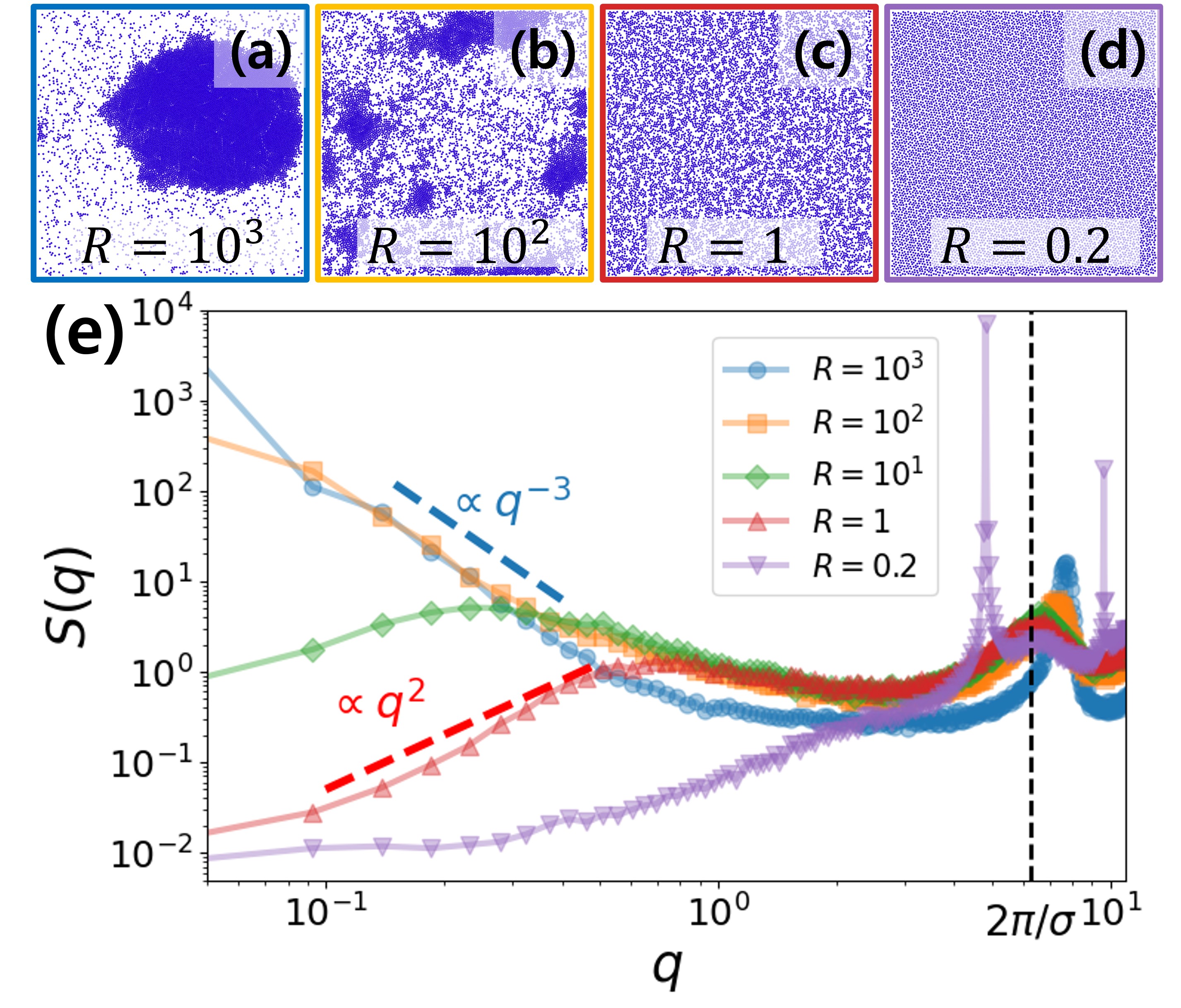}
  \caption{(a)--(d) Snapshots of the stationary state of cABP for
 various values of $R=v_0/\Omega$ at $\varphi=0.4$, $v_0=1$, and $\tau_p=200$.
(a) $R=10^3$, 
(b) $10^2$, 
(c) $1$, 
(d) $0.2$.  
(e) $S(\bm{q})$ for several $R$ values. 
The thick two broken lines indicate $q^{-3}$ and $q^2$. 
The thin vertical line shows the position of $q=2\pi/\sigma$. 
The system size is $N=10^4$.}
\label{fig1}
\end{figure}

\begin{figure}[tb]
\includegraphics[width=1.0\columnwidth,angle=-0]{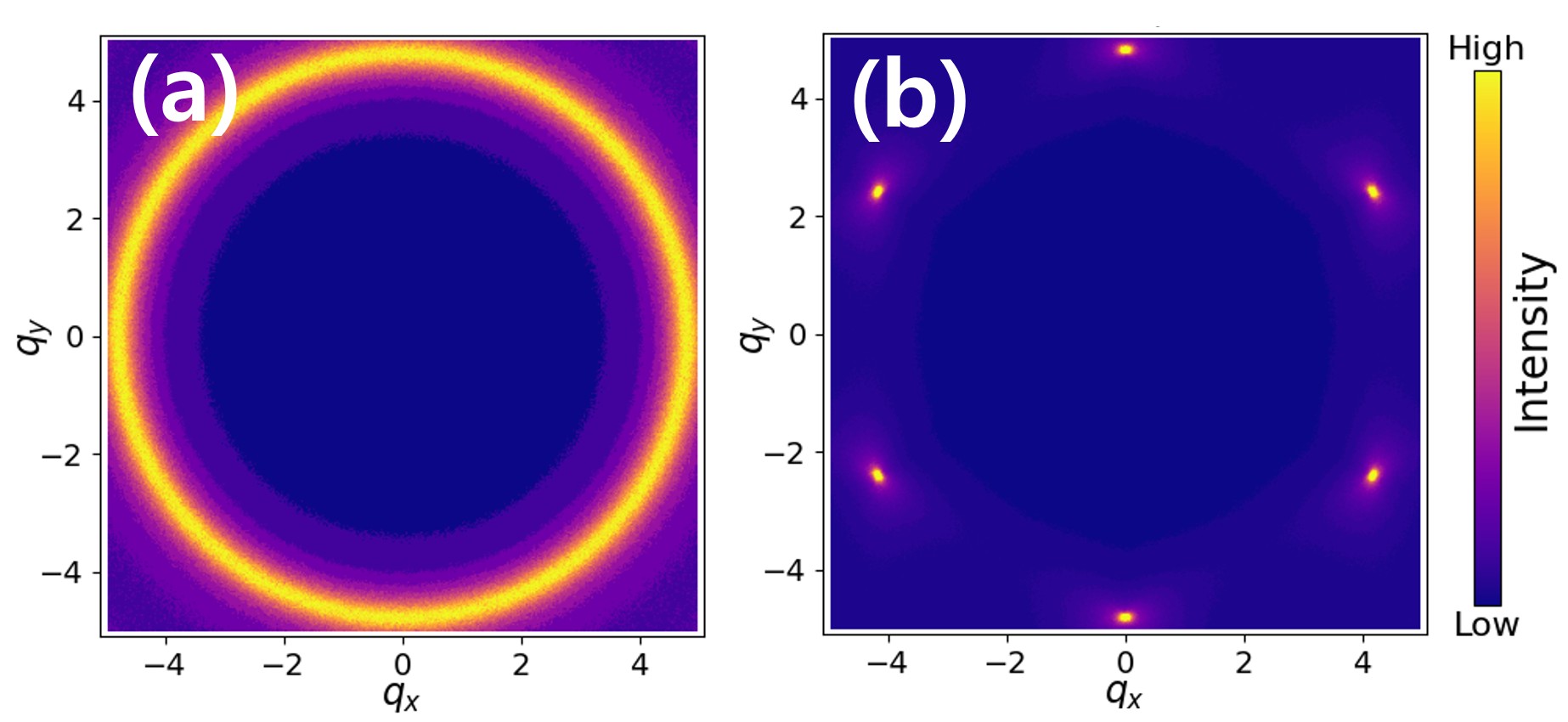}
\caption{
The heat map of the structure factor $S(\bm{q})$ for $\varphi=0.4$ and $\tau_p=10^2$ 
at (a) $R=0.299$ and (b) $R=0.294$. The color sidebar shows the intensity of the $S(\bm{q})$.
}
\label{fig2}
\end{figure}

\begin{figure}[tb]
\includegraphics[width=1.0\columnwidth]{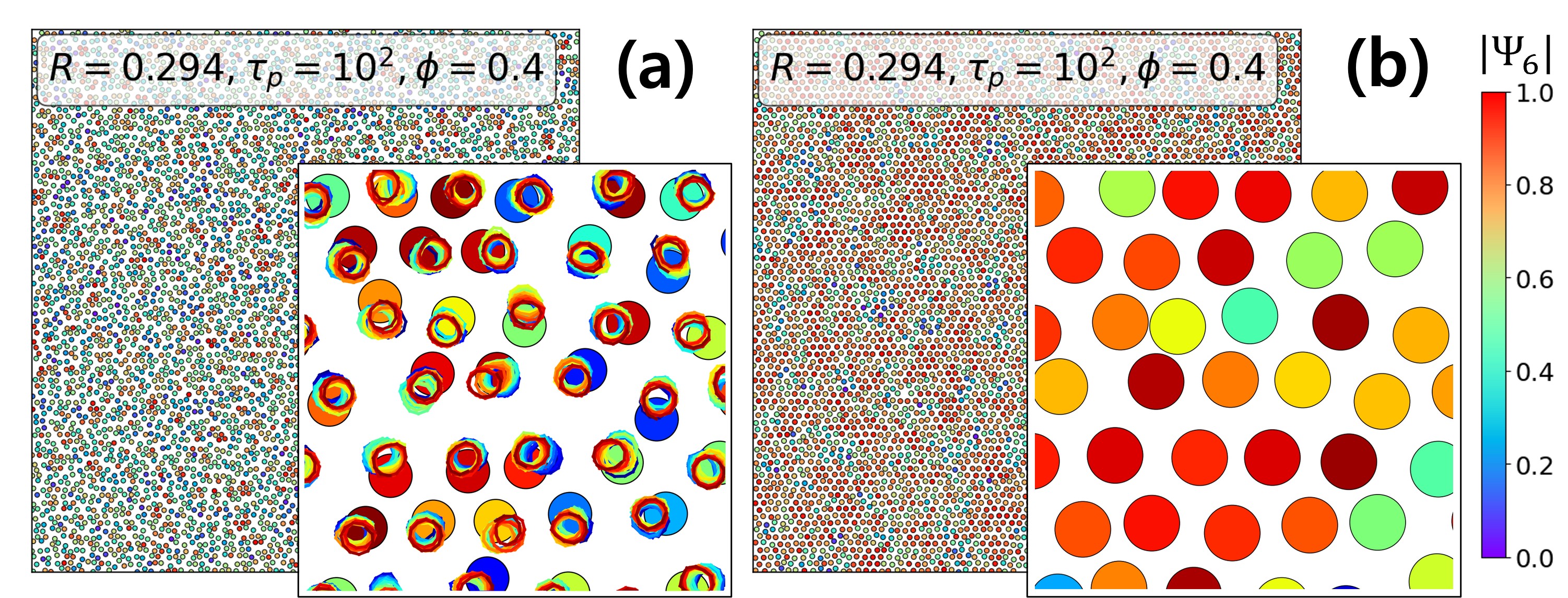}
\caption{(a) Snapshot of the particle configuration at $R=0.294$, $\tau_p=10^2$, and $\varphi=0.4$. 
The color of each particle represents the magnitude of its local hexatic
 order parameter $|\psi^i_6|$ (see the color bar).  
The inset shows a magnified view with particle trajectories superimposed to highlight their circular motion. 
(b) The configuration of the centers of mass of the orbits for the
 configuration in (a). 
The size of the circles representing the center positions is identical to that of the particles.}
\label{fig3}
\end{figure}

Unless otherwise stated, the packing fraction is set to
$\varphi=0.4$. 
This value is much smaller than the melting point for the equilibrium
hard discs $\varphi\approx0.7$ at which the system undergoes
the hexatic-solid transition~\cite{Bernard2011prl}. 
Throughout this work, $v_0$ is fixed to unity. 
The results remain qualitatively unchanged for other values of $v_0$~\cite{Lei2019sa}.
We first choose $ \tau_p = 200$, for which the P\'{e}clet number defined
by Pe$= \tau_p v_0 / \sigma $ is sufficiently large to induce
motility-induced phase separation (MIPS) when $R=\infty$ (or
$\Omega=0$)~\cite{Ma2022}. 
Figures~\ref{fig1} (a)--(d) show snapshots of particle configurations for several values of $R$. 
As expected, the system undergoes MIPS and forms large
clusters of densely packed regions when $R$ is large. 
As $R$ decreases, the clusters become smaller and the boundary of the phase-separated region becomes blurred~\cite{Lei2019sa}. 
At $R=0.2$, the system becomes spatially homogeneous.
Figure~\ref{fig1} (e) shows the static structure factor defined by
$S(\bm{q})=\ev{|\delta\rho_{\bq}|^2}/N$, where $\delta\rho_{\bq}$  is the
Fourier-transformed density fluctuation at wavevector $\bq$.
At $R=10^3$, where MIPS is clearly visible,
$S(\bm{q})$ diverges as $q^{-3}$ at small $q$, as predicted by Porod's law as a signature of phase separation~\cite{Onuki_2007,Kuroda2023}. 
As $R$ decreases below $R\approx10^2$, $S(\bm{q})$ at small $q$
decreases and exhibits power-law behavior $S(\bm{q})\sim q^\alpha$ with
a positive exponent $\alpha$. 
This is a clear sign of hyperuniformity~\cite{Torquato2018pr,Lei2019sa}.
Density fluctuations are suppressed because neighboring particles
constrain particle trajectories, which facilitates the dissolution
of large clusters.
This is consistent with the disappearance of MIPS observed in Fig.~\ref{fig1} (a)--(d).
It is theoretically established that the HU exponent $\alpha=2$ in the large $\tau_p$ limit~\cite{Lei2019sa,Kuroda2023jsmte}. 
HU behavior is observed around $R=1$ to $10$, though this algebraic behavior tapers off 
at small $q$. 
The crossover from HU to a constant $S(\bm{q})$ 
occurs when $\tau_p$ is finite~\cite{Lei2019sa}. 
Concomitantly, the first peak of $S(\bm{q})$ initially found around
$q=2\pi/\sigma$ shifts to lower $q$ and the width of the peak
widens. This reflects the fact that the tightly packed particles in the
MIPS clusters dissolve into a homogeneous fluid. 
As $R$ decreases below $2$, the MIPS clusters completely disappear and the
system becomes homogeneous. 
The height of $S(\bm{q})$ at small $q$ decreases and the HU behavior of $S(\bm{q})\sim q^2$ fades out. 
Surprisingly, at the smallest $R$ value of 0.2, the first peak suddenly sharpens and its height increases. 
This indicates the formation of crystals. 
Figure~\ref{fig2} shows the two-dimensional contour plot of $S(\bm{q})$
below and above the transition at which the sharp peak appears. 
The emergence of crystalline order with sixfold symmetry is clearly visible.
Figure~\ref{fig3} shows the instantaneous configurations of the
particles, which are colored according to the local hexatic order
parameter $|\psi_6^i|$ (see definition below).  
At first glance, crystalline order is not apparent in the configurations of the particles. 
They appear completely random. 
However, the crystalline order becomes clearly visible when the trajectories of particles are superimposed, as shown in the inset of Fig.~\ref{fig3} (a) 
(see also the movies available in the SM~\cite{SM}). 
The order becomes even more evident when one plots 
the positions of the centers of circular orbits, defined as
$\br_i^{0}\equiv \br_i-R\hat{\bm z}\times\hat{\bm e}_i$, as demonstrated
in Fig.~\ref{fig3} (b). 
Larger value of $|\psi_6^i|$ indicates the emergence of hexatic order.

\begin{figure}[tb]
\includegraphics[width=1.0\columnwidth]{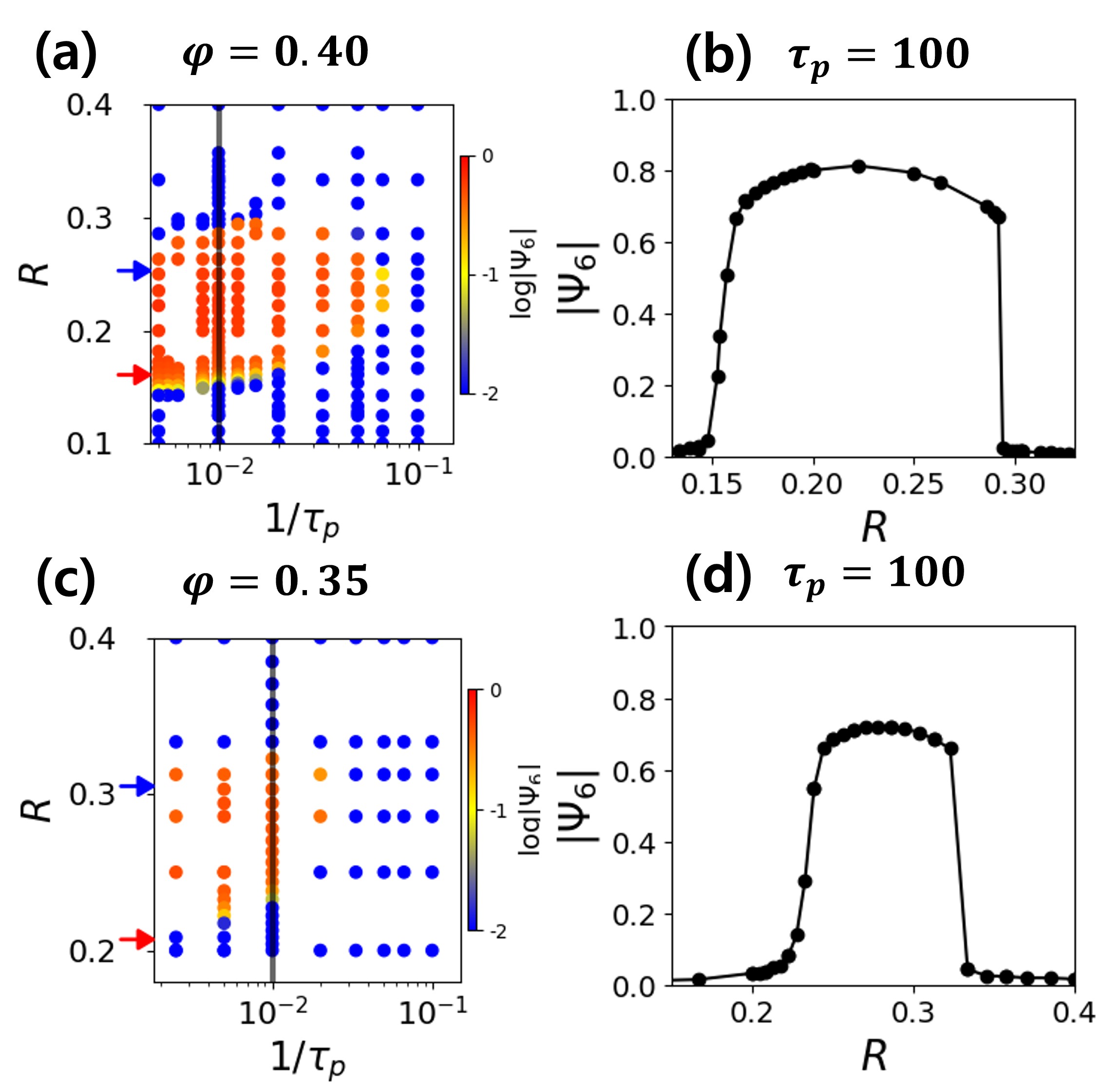}
\caption{
(a) Phase diagram at $\varphi = 0.4$. 
The color of the dots represents 
$|\Psi_6|$. 
Blue and red arrows indicate the upper and lower phase boundaries,
 respectively, as estimated from a geometric argument (see text).  
(b) The dependence of $|\Psi_6|$ on $R$ at $\tau_p = 10^2$ 
(shown as the solid line in (a)).
(c) and (d) present the same plots for $\varphi = 0.35$.
}
\label{fig4}
\end{figure}

In retrospect, the formation of the crystal is not surprising. A particle circulates with a radius $R$ and behaves like a larger particle with an effective radius $R_{\eff}= R+\sigma/2$, allowing the crystal to form at small
packing fractions, similar to a vortex lattice in superconductors~\cite{Guillamon2009natp}.
One would expect crystallization to occur when $R_{\eff}$ is small enough that all orbiting particles can be embedded in a hexatic lattice. 
At the same time, $R$ must be large enough so that the particles do
not melt into a fluid phase at low densities.  
Additionally, $\tau_p$ needs to be reasonably large 
for the crystal to form. 
Recall that, in the limit of $\tau_p\rightarrow 0$, the ABP model reduces to
an equilibrium Brownian particle system that is in a fluid phase at $\varphi=0.4$. 
Turning on $\Omega$ cannot change the phase diagram of the equilibrium
system, due to the Bohr-van Leeuwen theorem~\cite{Bohr1911,vanLeeuwen1919}. 
The subtle interplay of chirality ($\Omega\neq0$) and
being out of equilibrium ($\tau_p\neq0$) is essential for the crystallization.
We verify this prediction by simulating the system over a wide range of $R=v_0/\Omega$ and $\tau_p$. 
Figures~\ref{fig4} (a) and (c) show the phase diagram in the parameter space of $1/\tau_p$ and $R$ at $\varphi=0.40$ and $0.35$, respectively.
The largest values of $\tau_p$ explored are $200$ for
$\varphi=0.40$ and $10^3$ for $\varphi=0.35$, beyond which preparing stationary
states becomes prohibitively difficult.
The color of the dots in the phase diagrams represents the average of
the hexatic order parameter defined as
$\Psi_6 = N^{-1}\ev{\sum_{i=1}^{N} \psi_{6}^{i}}$,
where $\psi_{6}^{i}= N_{i}^{-1}\sum_{j}e^{6i\theta_{ij}}$ is the local
order parameter of the $i$-th particles. 
The summation is taken over the $N_i$ nearest neighbors of the $i$-th particle and
$\theta_{ij}$ is the relative angle of the bond between 
the center of the circular orbit of the particle and that of its nearest neighbors. 
The figures clearly demonstrate the main features discussed above:
the solid phase  in the finite range of $R$ confined by the upper and lower bounds of the binary lines, 
which eventually terminate at a small but finite $\tau_p$. 
Interestingly, the approximate locations of the lower and upper phase boundaries, $R_l$ and $R_u$, 
can be estimated by purely geometric argument.  
The lower boundary is estimated by equating 
the effective packing fraction of circulating particles, 
$\varphi_\text{eff} = \rho \pi R^2_{\text{eff}}$ (where $\rho$ is the number density), 
to the melting point of a hard-disk system, $\varphi_c \approx
0.7$~\cite{Bernard2011prl}, which leads 
\begin{equation}
R_{l} = \frac{\sigma}{2}\left( \sqrt{\frac{\varphi_c}{\varphi}} -1
			\right) \approx 0.16\sigma
\label{R_l}   
\end{equation}
for $\varphi=0.4$. 
Similarly, the upper boundary can be estimated to be $R_u \approx 0.25\sigma$
using the same equation by replacing $\varphi_c$ with the maximum hexatic packing 
$\varphi_m = \pi / (2\sqrt{3}) \approx 0.9$, 
assuming that particle collisions that would inevitably occur lead
to the breakdown of the hexatic order beyond this density.
These rough estimates are shown by the arrows in Fig.~\ref{fig4} (a) and (c), 
supporting our argument. 
For more quantitative discussion, an analysis based on the microscopic configurations would be required, 
which we leave for future work.
Note that the upper boundary $R_u$ 
is significantly lower 
than the
phase boundary of the absorbing phase transition observed in the
noiseless limit of $\tau_p=\infty$ at $R\approx1$ and
$\varphi=0.4$~\cite{Lei2019sa}. 
This indicates the possible existence
of a phase 
with configurational correlations or with
dynamical inhomogeneities between the active phase and the purely absorbing phase.

Recalling that the present system is two-dimensional, the next concern
is the nature of the fluid-solid transition that we are observing. On
the one hand, it is well-established that the two-dimensional
equilibrium fluids of particles with short-range interactions undergo a
two-step transition from the fluid phase to the hexatic phase to the
solid phase~\cite{Kosterlitz1973,Halperin1978,Young1979,Strandburg1988,Lee1992,
Zollweg1992, WeberMarxBinder1995, Jaster2004,Mak2006, Zahn1999prl, PengWangAlsayedYodhHan2010,
BagchiAndersenSwope1996}.  
This scenario has also been verified in several active matter models,
though the exponents characterizing the transition may
differ~\cite{Shi2023prl,Digregorio2018prl}. 
Conversely,
a recent study by Kuroda \textit{et al}. claimed 
otherwise for out-of-equilibrium systems and demonstrated that the
Mermin-Wagner fluctuations are strongly suppressed in cABP due to
hyperuniformity at high densities and in the large $\tau_p$ regime~\cite{Kuroda2025prr}.  
To determine which scenario applies to our system, 
we carefully examine the dependence of the order parameters and their
fluctuations.  
Figure~\ref{fig4} (b) shows the slice of the phase diagram at $\tau_p=10^2$
and $\varphi=0.4$, displaying the $R$ dependence of the modulus of the
global hexatic order parameter $|\Psi_6|$. 
Figure~\ref{fig4} (d) shows the same for $\varphi=0.35$.  
Two transitions are clearly observed but they are distinct.
At the first transition from a fluid to a 
hexatic phase around $R\approx0.15$, 
$|\Psi_6|$ sharply yet continuously increases,   
whereas $|\Psi_6|$ drops discontinuously at the second and re-entrant transition 
to a fluid phase around $R\approx0.3$.  
To characterize these transitions, we evaluate the two correlation functions. 
The first is the translational correlation function of the center of the
circular orbit which is defined by 
\begin{equation}
g_0(\br) \equiv\frac{1}{\rho N}\ev{\sum_{i,j}\delta(\br-\br^{0}_{ij})},
\label{gr}
\end{equation}
where $\br^{0}_{ij}=\br^{0}_i-\br^{0}_j$ and $\br_i^{0}$ is the instantaneous
position of the center of the circular orbit of the $i$-th particle.
Employing the centers of the circular orbits enhances 
the resolution of the correlations. 
The second is the orientational correlation function defined by
\begin{equation}
g_6(\br)\equiv \frac{1}{\rho N}\ev{\sum_{i,j}\psi^i_6\psi_6^{j*}\delta(\br-\br^{0}_{ij})}.
\label{g_6}
\end{equation}

\begin{figure}[tbp]
\includegraphics[width=1.0\columnwidth]{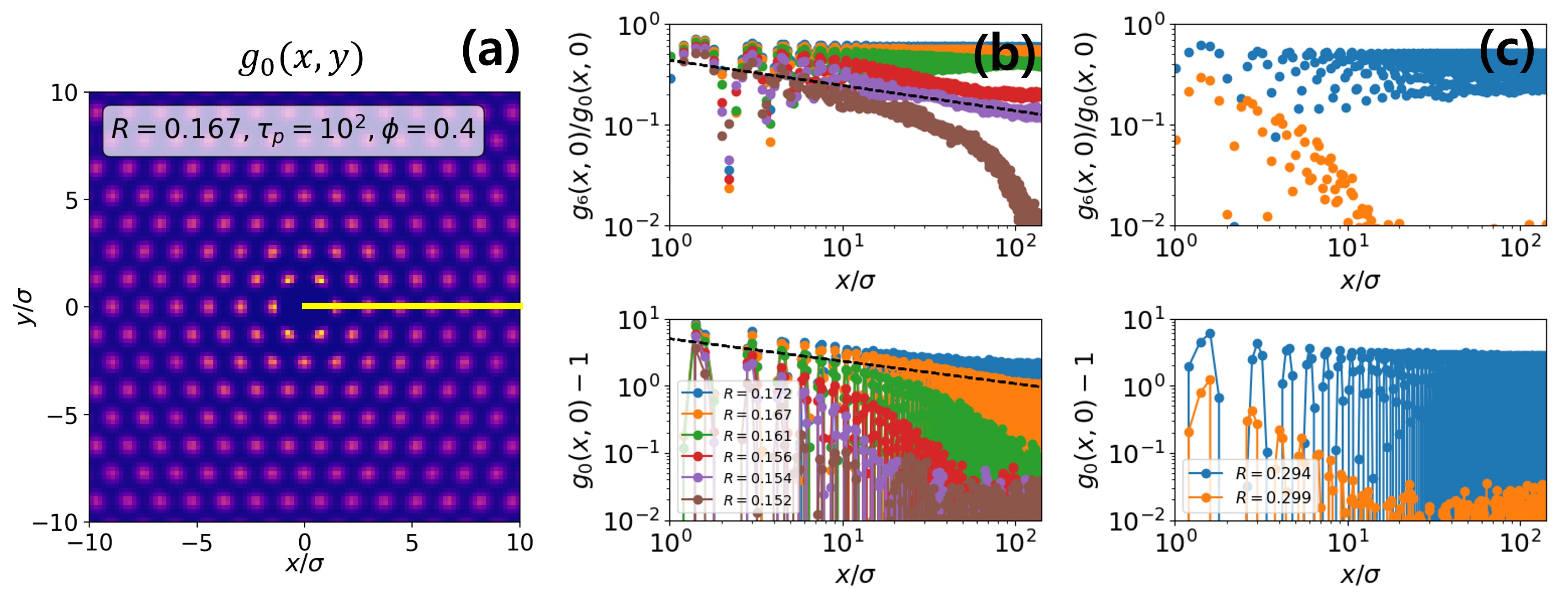}
\caption{
(a) $g_6(\br)/g_0(\br)$ 
and (b) $g_0(\br)-1$ (bottom)
along the direction of $y=0$ for several $R$ values near the first transition. 
(c), (d) Same as (a) and (b), but for $R$ near the second transition. 
The dashed lines in (a) and (b) represent the power laws 
$x^{-1/4}$
and 
$x^{-1/3}$, respectively. 
}
\label{fig5}
\end{figure}

Figure~\ref{fig5} (a) (b) presents one-dimensional cuts of 
the normalized orientational correlation $g_6(x, 0)/g_0(x, 0)$ 
and $g_0(x, 0)-1$ for various $R$ values 
between $0.152$ and $0.172$, in the vicinity of the first transition
shown in Fig.~\ref{fig4}(b).
The system size is $N = 40\,824$.
The two panels clearly demonstrate the characteristics of 
two-dimensional solidification predicted by the KTHNY scenario. 
At the low-$R$ side, the system is in a fluid phase, and both correlation
functions decay exponentially. 
As $R$ increases, $g_6(x,0)$ first
exhibits the algebraic decay $\sim x^{-\eta_6}$. 
The smallest $R$ at
which the power law is observed over the entire range of the windows of $x$
is $R_{\hex}\approx 0.154$ 
and the power law there is well fitted by using the
exponent $\eta_6=1/4$. 
This is the indication that the system undergoes the hexatic transition
at $R_\hex$. 
We also calculate the order parameter $\Psi_6$ for various system
sizes $N$. 
We find that $|\Psi_6|$ decays as $N^{-1/2}$ for $R < R_\hex$, a
trivial  $N$-dependence predicted for systems with short-ranged order.   
As expected for the quasi-long range ordered systems, 
$|\Psi_6| \sim N^{-\eta_6/4}$ at $R = R_\hex$ (see SM~\cite{SM}). 
As $R$ increases further, $\eta_6$ decreases and the $N$-dependence of $|\Psi_6|$
becomes weaker. 
Concomitantly, the translational correlation function $g_0(x,0)-1$
begins to develop a power law $\sim x^{-\eta_G}$.
The value at which the power law is seen over the entire range of the windows of $x$
is $R_G\approx0.162$, which is slightly larger than $R_\hex$. 
The corresponding exponent is close to $\eta_G=1/3$ as shown by the dotted line
in (b) 
and thus we conclude that $R_G$ marks the hexatic-solid transition. 
At larger $R$, $\eta_G$ keeps decreasing below 1/3. 
We also calculate the translational order parameter defined by 
$\rho_G=N^{-1}\ev{\sum_{j=1}^N e^{i\bG\cdot \br_j^{0}}}$,
where $\bG$ is the reciprocal lattice vector obtained from $S(\bq)$ in
Fig.~\ref{fig2} (b)~\cite{Li2019pre}. 
We find that $|\rho_G| \sim N^{-1/2}$ for $R < R_G$ and 
$\sim N^{-\eta_G/4}$ at $R = R_G$ \cite{SM}. 
For $R > R_G$, we find that $g_6(x,0)$ develops a plateau while
$g_0(x,0)-1$ continues to decay as $\sim r^{-\eta_G}$ 
where $\eta_G < 1/3$ and decreases with $R$.
Similarly, $|\rho_G|$ also decays as $N^{-\eta_G/4}$ with the same exponent 
as that of $g_0(x,0)-1$~\cite{SM}. 
These results suggest that the transition of circularly rotating
particles from a fluid to a hexatic crystal transition is two-step: 
The first is from the fluid phase to the hexatic phase in which the orientational order becomes quasi-long-range, whereas the translational order is short-ranged.
The second is from the hexatic to the solid phase, in which the orientational order becomes true long-ranged and the translational order becomes quasi-long-range. 
This is the same as the two-dimensional melting scenario in equilibrium 
systems~\cite{Kosterlitz1973,Lee1992,Halperin1978,Young1979}.   
We note that the nature of the transition that we observed 
at moderate densities and with finite $\tau_p$ in this study
is distinct from that at high densities and the large $\tau_p$ limit, where
the system develops the genuine long-range translational order and the
crystal phase is highly stable~\cite{Kuroda2025prr}.  

Finally, let us turn our attention to the re-entrant transition near $R_u$. 
As shown in Fig.~\ref{fig4} (c) and in {SM~\cite{SM}}, 
both $|\Psi_6|$ and $|\rho_G|$ drop discontinuously around $R\approx 0.28$. 
Figures~\ref{fig5} (c)--(d) show that neither $g_0(x,0)-1$ nor $g_6(x,0)$
exhibit a sign of algebraic decay. 
Of course, one cannot exclude the
possibility that the continuous and two-step transitions occur within an
extremely narrow range of $R$ values. 
However, we should not be surprised
even if the transition is discontinuous. If $R$ exceeds $R_u$ defined by
the maximum packing fraction of the effective particles, the orbiting
particles will inevitably collide. 
This collision destroys the circulating orbit and, consequently, the
effective particle picture no longer works {(see 
the video in SM~\cite{SM})}. 
This does not contradict the KTHNY scenario which does not consider 
the drastic changes to the internal degrees of freedom of the constituents (in this case, the destruction of
circular orbits).  

In this Letter, we have demonstrated the crystallization of the orbiting
active particles in two dimensions at low densities.
This is only possible if the system breaks the detailed balance of 
fluctuations and the chiral symmetry simultaneously.
The nature of the transitions at least at near $R_l$, or on the high $\Omega$-side, is
akin to an equilibrium two-step transition described by the KTHNY scenario. 
This scenario is consistent with the equilibrium systems of particles interacting with a soft potential~\cite{Kapfer2015prl}.
The next logical question concerns the fate of this transition line at
large densities. 
If both the density and $\tau_p$ are large enough, the scenario is
completely different, and the hyperuniformity and packing effects
stabilize the crystal~\cite{Kuroda2025prr}. 
Finally, we note that the largest system size of $N = 40\,824$ explored in this study is 
still moderate for determining whether the fluid-hexatic transition 
is continuous or not, as corroborated by the large-scale simulations of monatomic equilibrium systems~\cite{Bernard2011prl,Kapfer2015prl}.
A more elaborate and systematic analysis using the larger systems is
left for future work.

\begin{acknowledgments}
We thank Yoshihiko Nishikawa and Harukuni Ikeda 
for their insightful comments and fruitful discussion.
This work was supported by KAKENHI (Grant Numbers 
JP20H00128, 
JP22H04472, 
JP23H04503, 
JP23KJ1068, 
JP24H00192),  
and the JST FOREST Program (Grant Number JPMJFR212T). 
\end{acknowledgments}

\begin{center}
{\bf DATA AVAILABILITY}
\end{center}
The data that support the findings of this study are available
from the corresponding author upon reasonable request.

%

\end{document}